\documentclass[epj]{svjour}
\usepackage{amsmath,amssymb}
\usepackage{times}
\usepackage{txfonts}
\usepackage{dcolumn}
\usepackage{bm}
\usepackage{lineno}
\usepackage{graphicx}

\newcommand{\del}{\mathrm{d}}

\usepackage{lscape}

\usepackage{longtable}
\DeclareSymbolFont{EulerExtension}{U}{euex}{m}{n}
\DeclareMathSymbol{\euintop}{\mathop} {EulerExtension}{"52}
\DeclareMathSymbol{\euointop}{\mathop} {EulerExtension}{"48}

\everymath{\displaystyle}
\begin{document}
\titlerunning{
Modernization in processing and dissemination of experimental cross section ...
}
\title{
Modernization in processing and dissemination of experimental cross section data in EXFOR
}

\authorrunning{Naohiko Otuka, Vidya Devi}
\author{
Naohiko Otuka\inst{1}\thanks{n.otsuka@iaea.org}
\and
Vidya Devi\inst{1}
}
\institute{
Nuclear Data Section,
Division of Physical and Chemical Sciences,
Department of Nuclear Sciences and Applications,
International Atomic Energy Agency,
A-1400 Wien,
Austria
}

\date{Received: date / Revised version: date}

\abstract{
The EXFOR library is a nearly complete database for experimental radionuclide production cross sections for light charged-particle induced reactions,
and retrieval of the experimental cross sections in EXFOR is the first step in determination of the best estimate of the radionuclide production cross section.
The EXFOR format was designed with 80-column punched cards as the recording media,
but the format is still preferred for compilation and exchange of the nuclear reaction data.
The format is suitable for reading and processing by EXFOR compilers,
but it is not a good choice for reading by EXFOR users with modern computer languages such as Python.
We have recently attempted to lower the barrier for modern software engineers to access the information stored in the EXFOR by developing tools converting an EXFOR file to (1) the HTML format for human reading (X4IVEW), and (2) the CX4 (Compact EXFOR) format for tabulation and plotting (X4TOCX).
This article introduces X4VIEW and a preliminary version of X4TOCX with the radionuclide production cross section as an example.
}

\maketitle
\onecolumn
\section{Introduction}
\label{sec:introduction}
Production of radionuclides for medical application (diagnosis and therapy) is being routinely done by irradiation of materials by charged-particle beams extracted from cyclotrons.
Daily production of radionuclides requires optimization of the production route maximizing the activity of the desired radionuclide and, at the same time, minimizing the activities of the undesired by-product radionuclides.
The end-of-bombardment (EOB) thick target yield (activity at EOB per beam current~\cite{Otuka2015}) of a radionuclide after irradiation of a target (density $\rho$) for time $t$ is 
\begin{equation}
a(t)=\frac{1-e^{-\lambda t}}{\lambda}\int_{E_0}^{E_1}\del E \frac{\sigma(E)}{S(E)},
\end{equation}
where $\lambda$ is the decay constant of the radionuclide, $E_0$ and $E_1$ are the beam particle energies at the exit and entrance of the target ($E_0$=0 if the initial beam energy $E_1$ is fully absorbed by the target), $\sigma(E)$ is the radionuclide production cross section and $S(E)=-\del E/\del (\rho x)$ is the stopping power of the material ($x$: path length of the beam particle).
This equation demonstrates that the energy dependent radionuclide production cross section (excitation function) is an essential parameter in estimation of the radionuclide activity earned by the irradiation.

The IAEA Nuclear Data Section (NDS) has recognized importance of medical radionuclide production cross sections since the 1980s~\cite{Okamoto1988}.
The Coordinated Research Project (CRP) on ``Charged particle cross-section database for medical radioisotope production: Diagnostic radioisotopes and monitor reactions" (Project\# F41014, 1995-1999)~\cite{IAEA2001} released a database of the cross sections recommended by the experts,
and the database was updated and extended in a recent CRP on ``Nuclear data for charged-particle monitor reactions and medical isotope production"(Project\# F41029, 2012-2018)~\cite{Engle2019,Tarkanyi2019a,Tarkanyi2019b,Hermanne2021,Tarkanyi2022,Hermanne2023a,Hermanne2023b,Tarkanyi2024,Hermanne2025}.
These projects determined the recommended cross sections by fitting a function to the experimental data points without nuclear physics modelling.
Namely, this approach requires collection of experimental data at the beginning.

The EXFOR library~\cite{Otuka2014} is a collection of experimental reaction data widely used in nuclear physics and engineering.
The EXFOR project was initially for low-energy neutron energy application,
and there was no significant attempt to maintain EXFOR as a complete database for charged-particle induced reaction cross sections until c.a. 2000.
Consequently,
the first (1995-1999) CRP had to visit not only EXFOR but also other data collections such as Landolt-B\"ornstein Series~\cite{LB} and original publications.
After this CRP,
the International Network of Nuclear Reaction Data Centres (NRDC) made a significant effort to improve completeness of the EXFOR library for radionuclide production applications by comparing the data compiled in the EXFOR with those in Landolt-B\"ornstein compilation as well as Nuclear Science Reference (NSR) database~\cite{Pritychenko2011},
and now we believe EXFOR is nearly complete for the radionuclide production cross sections for light charged-particle (proton, deuteron, triton, helion and alpha-particle) induced reactions.

The EXFOR library is a set of EXFOR entries,
and each EXFOR entry provides numerical data and their description (e.g., quantity definition, experimental instruments) of an experimental work in the EXFOR format,
which was designed in the 1970s for storing the information in 80-column punched cards (Hollerith cards) and for reading by Fortran.
Various tools handling files in the EXFOR format (e.g., EXFOR Editor~\cite{Pikulina2024}, JANIS Trans Checker~\cite{Soppera2017}) are available,
and use of the format does not make a bottleneck in EXFOR compilation and exchange process for EXFOR developers.
However,
development of codes reading an EXFOR file by software engineers has been challenging since they need to develop parsers interpreting the characters in the 80-column format.
Conversion of EXFOR files to XML (Extensible Markup Language) files was discussed as a solution to overcome the problem,
and Zerkin adopted XML as an intermediate output for construction of an SQL database for the IAEA NDS EXFOR web retrieval system~\cite{Forrest2014}.
Afterwards,
Schnabel pointed out that JSON (JavaScript Object Notation) is another format widely supported in the field of information technology and is a natural candidate for conversion of EXFOR files~\cite{Schnabel2020}.
Motivated by his idea,
we designed a JSON representation of EXFOR (J4),
and started distribution of J4 files synchronized with the latest EXFOR library by continuous conversion of the latest EXFOR files by X4TOJ4, a newly developed EXFOR-to-J4 converter included in the ForEXy code package~\cite{Otuka2025} and distributed from the PyPI repository~\cite{PyPI2026}.
To examine its usefulness,
we developed a Python tool for construction of experimental cross section covariances from J4 files,
and applied them to evaluation of the $^{237}$Np fast neutron fission cross section by the simultaneous least-squares fitting~\cite{Devi2025}.

The radionuclide production cross sections recommended by evaluators are also often determined by least-squares fitting to the experimental cross sections (e.g.,~\cite{Hermanne2018}).
Preparation of the experimental inputs to fitting tools is typically done by copying and pasting the EXFOR source file to a spreadsheet with manual editing, which is not adequate from the view of both accuracy and efficiency.
This situation motivated us to develop a modern pipeline generating files adopting a machine-readable representation of the EXFOR library.

In connection with our recent works on radioisotope production cross section measurements~\cite{Otuka2024,Otuka2025b},
we noticed that there are two more applications that could be incorporated into the newly developed EXFOR processing pipeline powered by J4.
They are generation of (1) a human readable representation of the EXFOR information, and (2) a compact plotting-oriented representation of the EXFOR data.

There have been attempts to provide such representations as a part of nuclear data services.
For the human readable representation, Zerkin et al. developed the X4+ and X4$\pm$ representations~\cite{Zerkin2018} to make some cryptic EXFOR abbreviations human readable (e.g., expansion of \texttt{JRN} to ``Journal of Radioanalytical and Nuclear Chemistry").
However,
these representations do not support some important keywords such as decay data (e.g., \texttt{(78-PT-195-M,4.010D,DG,98.9,0.117)} for 98.9~keV gamma line ($I_\gamma$=11.7\%) of $^{195m}$Pt with $T_{1/2}$=4.010~d),
and it has been desired to have a human readable representation supporting all keywords of EXFOR.
For the plotting-oriented representation, the C4/C5 representations~\cite{Zerkin2018,Cullen2001} have been widely used for many decades.
However,
these representations pack full information of each data point in a single line,
and there have been attempts to develop more user-friendly table representation (e.g., EXFORTABLE~\cite{Koning2015}.
Under these situations,
it would be meaningful to extend our EXFOR processing pipeline for generations of human readable EXFOR representation and compact plotting-oriented EXFOR representation.

This article first briefly summarizes typical use of EXFOR in evaluation of radionuclide production cross sections,
and then introduces development of EXFOR processing tools for better readability for humans (X4VIEW) and for tabulation and plotting (X4TOCX).

\section{Use of EXFOR in radionuclide production cross section evaluation}
\label{sec:evaluation}
The radionuclide $^{99m}$Tc ($T_{1/2}$=6.0066~h)~\cite{Kondev2021} is widely used as diagnostic nuclear medicine due to the 140.5~keV $\gamma$-ray emitted in its isomeric transition to the ground state,
and it constitutes more than 70\% of the medical imaging procedures~\cite{Tarkanyi2019b}.
For this purpose,
$^{99m}$Tc has been extracted from the $^{99}$Mo/$^{99m}$Tc generator with $^{99}$Mo produced in $^{235}$U thermal neutron fissions in reactors in these decades.
It is uncertain if this production route is sustainable due to aging of the reactors providing $^{99}$Mo,
and alternative routes for production of $^{99}$Mo or $^{99m}$Tc by charged-particle induced reactions (e.g., $^{100}$Mo(p,x)$^{99}$Mo, $^{100}$Mo(p,2n)$^{99m}$Tc) and photonuclear reactions (e.g., $^{100}$Mo($\gamma$,n)$^{99}$Mo) are actively being studied~\cite{IAEA2017}.
Here,
we briefly demonstrate how evaluators get experimental cross sections and utilize them with $^{100}$Mo(p,2n)$^{99m}$Tc as an example.

\begin{figure}[hbtp]
\begin{center}
\includegraphics[width=0.46\textwidth,bb=0.000000 0.000000 469.565544 416.308110]{"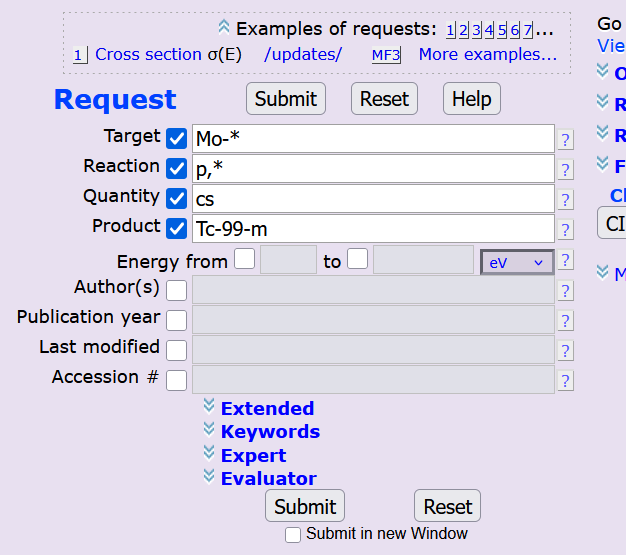"}
\includegraphics[width=0.43\textwidth,bb=0.000000 0.000000 413.307691 393.054864]{"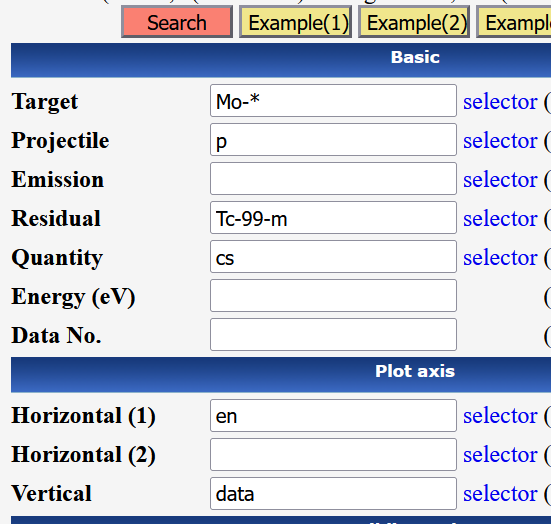"}
\caption{
Retrieval of Mo(p,x)$^{99m}$Tc cross sections on EXFOR web retrieval systems maintained by NDS (left) and JCPRG (right).
\texttt{Mo-*} is used to find both $^{100}$Mo(p,2n)$^{99m}$Tc and $^\mathrm{nat}$Mo(p,x)$^{99}$Mo cross sections (* is a wildcard).
}
\label{fig:exforweb}
\end{center}
\end{figure}

Table~\ref{tab:exforweb} summarizes the EXFOR web retrieval systems with indication of the last updated date and maintained by the NRDC members.
These web retrieval systems return a list of datasets satisfying the request,
and users can plot them on the systems.
Figure~\ref{fig:exforweb} shows how to submit a request for Mo(p,x)$^{99m}$Tc production cross sections on the NDS and JCPRG web retrieval systems.
The plots generated for such a request exhibits the presence of two groups of datasets having the same energy dependence but with an order of magnitude difference.
This is because only $^{100}$Mo is a target isotope contributing to $^{99m}$Tc production,
and the experimentalists can publish the cross sections as either
(1) $^\mathrm{nat}$Mo(p,x)$^{99m}$Tc \textit{elemental} cross sections (\texttt{42-MO-0(P,X)43-TC-99-M} in EXFOR) or
(2) $^{100}$Mo(p,2n)$^{99m}$Tc \textit{isotopic} cross sections (\texttt{42-MO-100(P,2N)43-TC-99-M} in EXFOR).
Therefore,
the evaluators cannot get an overview of the existing datasets by plotting on these web systems.
Instead of this quick plotting,
the users have to download the datasets from the web retrieval systems and divide the elemental cross section datasets in EXFOR by the natural isotopic abundance of $^{100}$Mo ($\sim$0.097~\cite{Meija2016}) for comparison with the isotopic cross section datasets in EXFOR,
or vice versa.

\begin{table}[hbtp]
\begin{center}
\caption{
EXFOR web retrieval systems with indication of the last updated dates (as of 1 June 2026).
}
\label{tab:exforweb}
\begin{tabular}{lll}
\hline
URL                               &Centre           & Last update \\
\hline                           
https://nds.iaea.org/exfor/       &NDS (Vienna)     & 2026-05-28  \\
https://www.jcprg.org/exfor/      &JCPRG (Sapporo)  & 2026-05-19  \\
https://nds.iaea.org/dataexplorer/&NDS (Vienna)     & 2023-10-27  \\
http://cdfe.sinp.msu.ru/exfor/    &CDFE (Moscow)    & 2019-05-11  \\
\hline
\end{tabular}
\end{center}
\end{table}

The users also sometimes need to update a dataset with the latest reference values.
The radionuclide production cross section $\sigma_x$ measured relative to the beam flux monitor reaction cross section $\sigma_m$ by the decay gamma spectroscopy is related to the counts in the gamma peak corrected for the detection efficiency $C$ by
\begin{equation}
\sigma_x=\frac{\lambda_x C_x}
              {\lambda_m C_m}
         \frac{n_mI_m(1-e^{-\lambda_m t_i})e^{-\lambda_m t_c}(1-e^{-\lambda_m t_m})}
              {n_xI_x(1-e^{-\lambda_x t_i})e^{-\lambda_x t_c}(1-e^{-\lambda_x t_m})}\sigma_m,
\end{equation}
where $n$ is the areal density of target atoms, $I$ is the decay gamma intensity, $t_i$, $t_c$ and $t_m$ are the irradiation time, cooling time and counting time.
The three parameters on the right-hand side, $I_m$, $I_x$ and $\sigma_m$, are usually taken from nuclear databases (e.g., $I$ from IAEA LiveChart of Nuclei~\cite{Verpelli2011} and $\sigma_m$ from the IAEA Reference Cross Sections for Charged-Particle Monitor Reactions~\cite{Hermanne2018}).

\begin{figure}[hbtp]
\centering
\includegraphics[bb=0 0 595 842,angle=0,width=0.6\linewidth]{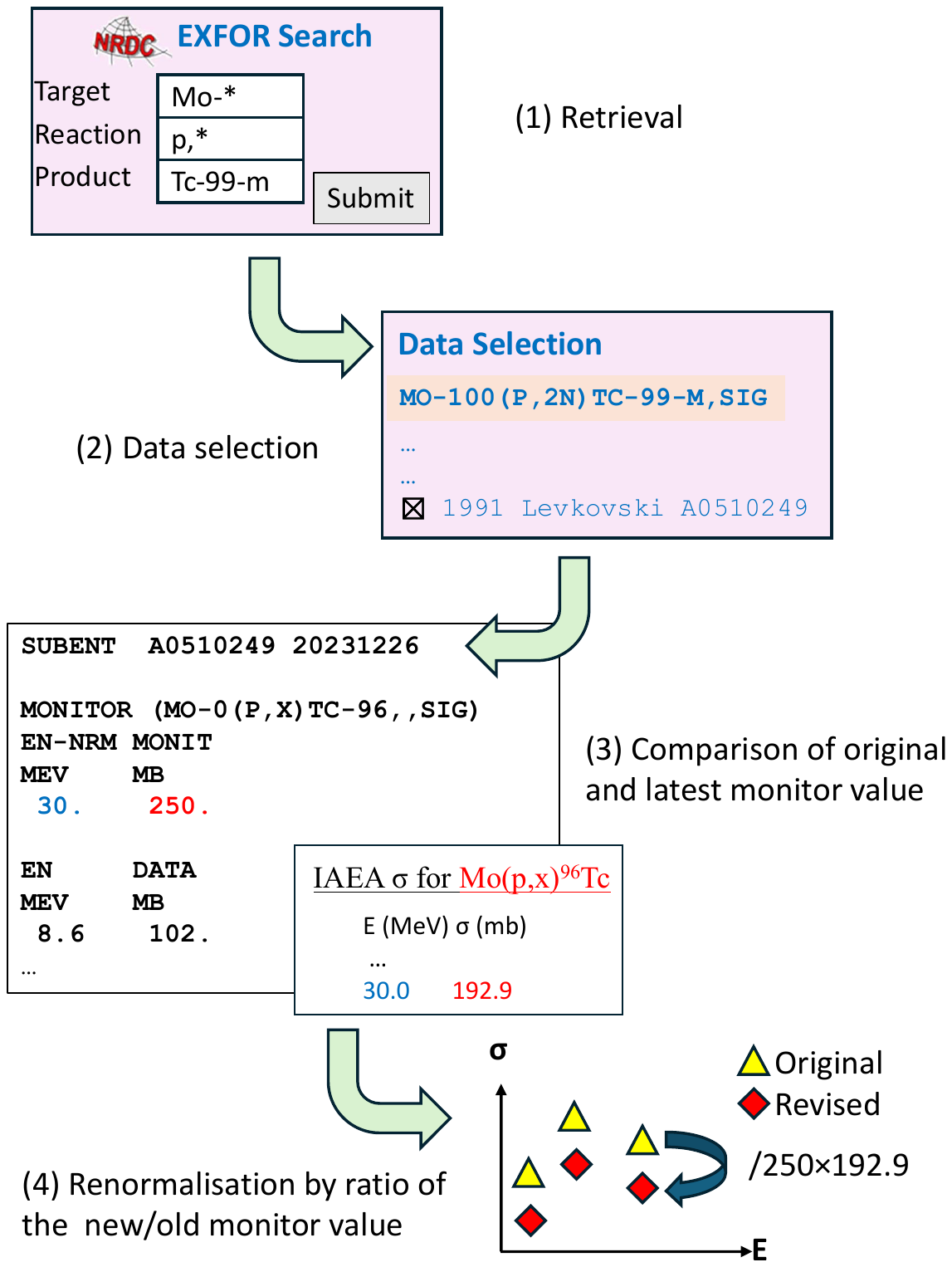}
\caption{
Flow of renormalization for the $^{100}$Mo(p,2n)$^{99m}$Tc cross section due to change of the $^\mathrm{nat}$Mo(p,x)$^{96}$Tc monitor cross section from 250~mb (the value adopted by the experimentalists~\cite{Levkovskij1991}) to 192.9~mb (the latest recommended value~\cite{Hermanne2018}).
}
\label{fig:renormalization}
\end{figure}

These reference values are continuously updated,
and cross sections determined with outdated reference values must be renormalized by using the latest reference values.
For this purpose,
the cross section evaluators should look for the reference values adopted by the experimentalist in the EXFOR file.
Among the datasets found by the retrieval shown in Fig.~\ref{fig:exforweb}, 
for example,
the $^{100}$Mo(p,2n)$^{99m}$Tc cross section measured by Levkovskij~\cite{Levkovskij1991} is compiled in EXFOR A0510.250 with the $^\mathrm{nat}$Mo(p,x)$^{96}$Tc cross section of 250$\pm$10~mb at 30~MeV as $\sigma_m$.
The corresponding value recommended by the IAEA in 2018~\cite{Hermanne2018} is 192.9$\pm$8.4~mb,
and hence the cross section compiled in EXFOR A0510.250 from the original publication must be updated with the renormalization factor of 192.9/250$\sim$0.77~\cite{Hermanne2018}.
Figure~\ref{fig:renormalization} shows the workflow for renormalization of the cross sections compiled in EXFOR due to the revision of the monitor cross section.

\begin{figure}[hbtp]
\begin{center}
\includegraphics[width=0.49\textwidth,bb=0.000000 0.000000 578.330725 353.299315]{"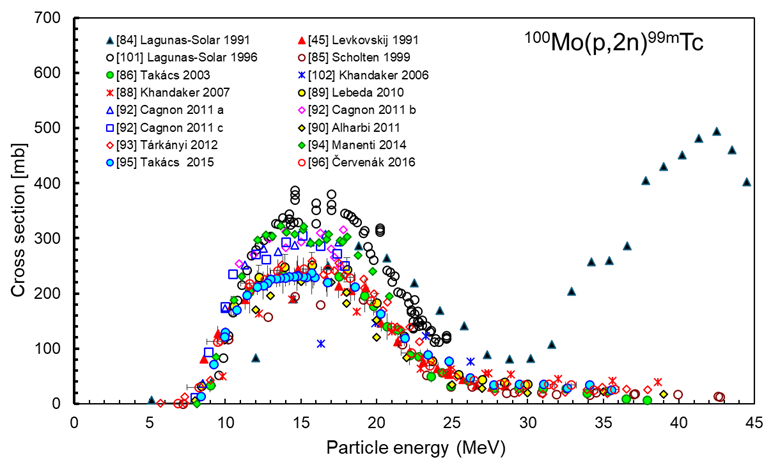"}
\includegraphics[width=0.49\textwidth,bb=0.000000 0.000000 578.330725 352.549210]{"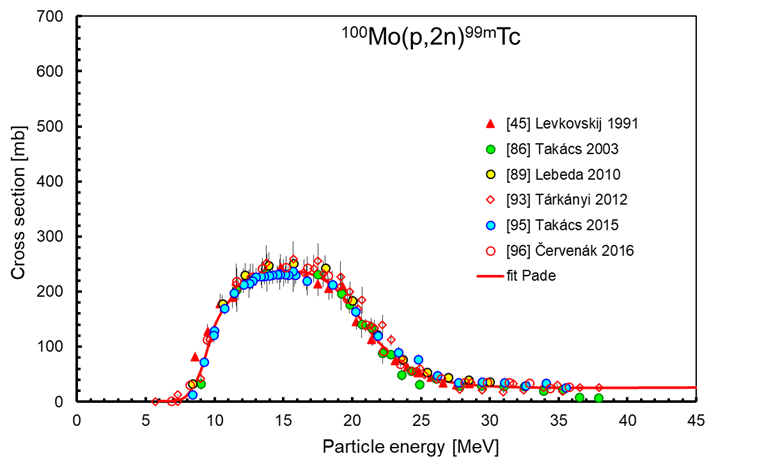"}
\caption{
$^{100}$Mo(p,2n)$^{99m}$Tc cross section datasets extracted from the EXFOR web retrieval systems and corrected for the known effects: All datasets (left) and the datasets selected for evaluation (right).
Images courtesy of S\'{a}ndor Tak\'{a}cs (ATOMKI).
}
\label{fig:100Mopx99mTc}
\end{center}
\end{figure}

The left panel of Fig.~\ref{fig:100Mopx99mTc} shows a plot of the $^{100}$Mo(p,2n)$^{99m}$Tc production cross sections renormalized for the known effects.
This figure shows that the existing experimental datasets exhibit large discrepancy even after the renormalization.
Tak\'acs et al. discusses this discrepancy in the relation with the complex nature of the 140.5~keV gamma emission~\cite{Takacs2016},
which is the only usable gamma line for quantification of $^{99m}$Tc but emitted 
\begin{itemize}
\item in isomeric transition of $^{99m}$Tc produced directly in $^{100}$Mo(p,2n)$^{99m}$Tc reaction
\item in isomeric transition of $^{99m}$Tc produced indirectly in $^{100}$Mo(p,x)$^{99}$Mo $\to$ $^{99m}$Tc
\item following $\beta^-$ decay of $^{99}$Mo produced directly in $^{100}$Mo(p,x)$^{99}$Mo reaction.
\end{itemize}
The evaluators~\cite{Tarkanyi2019b} selected reliable datasets from those in Fig.~\ref{fig:100Mopx99mTc} (left),
and applied the Pad\'e fit to the selected datasets as shown in Fig.~\ref{fig:100Mopx99mTc} (right).

The story on the renormalization and selection of the $^{100}$Mo(p,x)$^{99m}$Tc experimental cross sections shows that the EXFOR users should be able to catch the key information (e.g., reference values) in the EXFOR file, and also to manipulate the numerical data stored in the EXFOR format.
In the following two sections,
we introduce two new EXFOR processing tools helping the EXFOR users performing these tasks.

\section{X4VIEW: Converter for reading EXFOR files}
\label{sec:x4view}
It is not always difficult to read an EXFOR file.
Apart from the data table records,
REACTION is the most important record keyword as it defines the quantity considered.
One can easily infer that the REACTION record of the EXFOR entry shown in Fig.~\ref{fig:100Mopx99MoX4} expresses the $^{100}$Mo(p,x)$^{99}$Mo cross section,
and could infer that \texttt{(CUM),SIG} indicates something relevant to the cumulative cross section (i.e., cross section for $^{99}$Mo production including contribution of $^{99}$Nb $\beta^-$ decay).
Proper understanding of REACTION is sufficient for those who just want to plot the dataset.
However,
evaluators often need more details.
If the dataset shows deviation from other datasets,
they probably want to find the reference values of the decay gamma intensities and beam monitor cross section adopted by the experimentalist to seek possible renormalization for better agreement with other datasets.
However, the numbers coded under the keyword DECAY-DATA in Fig.~\ref{fig:100Mopx99MoX4} look cryptic,
and probably they should ask a data center for help, or read the EXFOR Formats Manual~\cite{Otuka2025a} to understand the meaning of each number under DECAY-DATA.
\begin{figure}[hbtp]
\begin{center}
\begin{minipage}[c]{0.80\textwidth}
\begin{small}
\begin{verbatim}
SUBENT        C2156002   20150310                             C148C215600200001
BIB                  8         17                                 C215600200002
REACTION   (42-MO-100(P,X)42-MO-99,(CUM),SIG)                     C215600200003
DECAY-DATA (42-MO-99,65.94HR,DG,140.511,0.0452,                   C215600200004
            DG,181.068,0.0599,DG,739.500,0.1213)                  C215600200005
PART-DET   (DG)                                                   C215600200006
SAMPLE     (42-MO-100,ENR=0.9742)                                 C215600200007
...
COMMON               4          3                                 C215600200021
ERR-1      ERR-2      ERR-3      ERR-4                            C215600200022
PER-CENT   PER-CENT   PER-CENT   PER-CENT                         C215600200023
 5.0        1.0        5.0        5.0                             C215600200024
ENDCOMMON            3          0                                 C215600200025
DATA                 4          8                                 C215600200026
EN         EN-RSL     DATA       ERR-T                            C215600200027
MEV        MEV        MB         MB                               C215600200028
 10.0       0.1        0.264      0.036                           C215600200029
 11.3       0.1        3.140      0.280                           C215600200030
 12.5       0.1        9.890      0.870                           C215600200031
...
\end{verbatim}
\end{small}
\end{minipage}
\caption{EXFOR file providing a $^{100}$Mo(p,x)$^{99}$Mo cross section dataset (uncertain if cumulative).}
\label{fig:100Mopx99MoX4}
\end{center}
\end{figure}

X4VIEW is a Python module converting an EXFOR file to a human readable HTML file.
This script converts an EXFOR file to an HTML file by calling three tools: X4TOJ4 (EXFOR-to-J4 converter), POIPOI (post-processing of J4 to remove ``pointers" in EXFOR)\footnote{
See Section~\ref{sec:x4tocx} for the concept of the pointer and the role of POIPOI.
}
and J4VIEW (J4-to-HTML converter).
See Ref.~\cite{Otuka2025} for details about X4TOJ4 and POIPOI.
 
Conversion of an EXFOR file \textit{exfor.txt} to an HTML file \textit{exfor.html} for the EXFOR entry C2156 can be done by
\begin{verbatim}
x4_x4view.py -i exfor.txt -e c2156 -o exfor.html
\end{verbatim}
.
X4VIEW supports all EXFOR keywords and preserves all pieces of the information in the EXFOR file.
Users can customize the style of the HTML outputs by modifying \textit{exfor.css}.
If a reference has a DOI and option (\texttt{-m}) to insert the DOI is activated, a hyperlink is added to the reference in the HTML output.

\begin{figure}[hbtp]
\begin{center}
\includegraphics[width=0.55\textwidth,bb=0.000000 0.000000 591.989262 562.727436]{"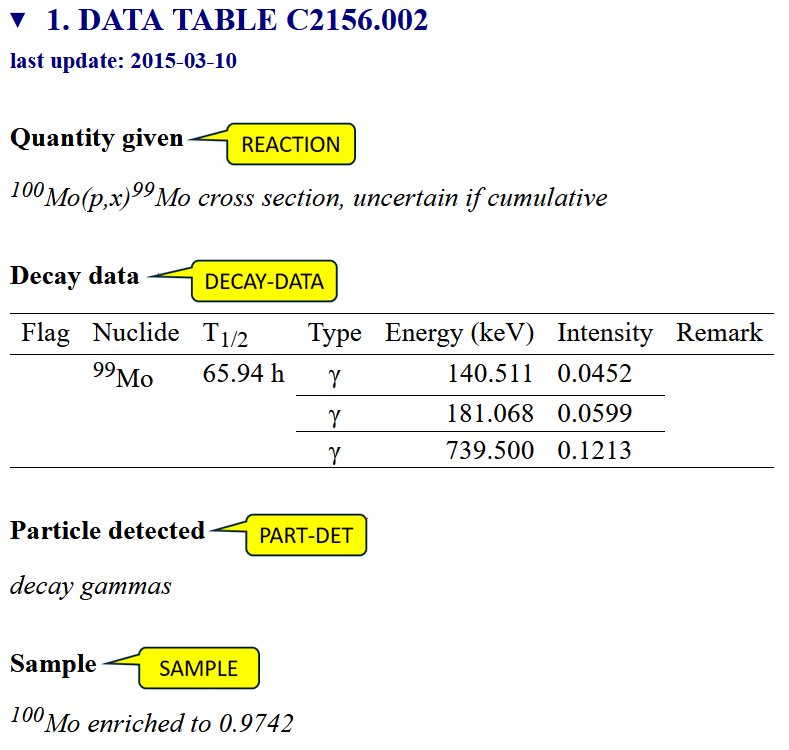"}
\includegraphics[width=0.40\textwidth,bb=0.000000 0.000000 367.648591 366.898288]{"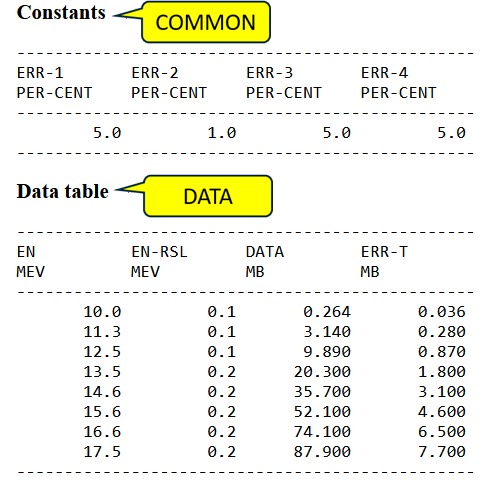"}
\caption{
HTML representation of EXFOR C2156.002 (excerption): Data description part (left) and numerical data part (right).
The callouts show the corresponding keywords in the EXFOR file (Fig.~\ref{fig:100Mopx99MoX4}).
Note that the COMMON section part is omitted in Fig.~\ref{fig:100Mopx99MoX4}.
}
\label{fig:x4view}
\end{center}
\end{figure}

Figure~\ref{fig:x4view} shows an excerption from the HTML output of EXFOR C2156.002 generated by X4VIEW,
which clearly shows that the numbers listed under DECAY-DATA are the gamma energies and gamma intensities.
We found this HTML output is very useful not only for EXFOR users but also for proofreading of an EXFOR entry by the authors.
Authors are usually not familiar with the EXFOR format,
and the compilers cannot be sure to what extent the authors understood the EXFOR file even they approve release of the EXFOR file.
Contrarily,
an EXFOR file converted to HTML is self-explanatory,
and we can expect more meaningful proof by the authors.
Now the IAEA NDS sends an EXFOR file draft to the authors together with its HTML output (printed in a pdf file) and receives from authors the proof pdf files annotated by their comments.
Namely,
the HTML output improves the quality of the EXFOR file by deepening the interaction between the compiler and authors.
X4VIEW was released in May 2026 as a part of the ForEXy code package distribution~\cite{PyPI2026}.

\begin{figure}[hbtp]
\centering
\includegraphics[bb=0 0 800 600,angle=0,width=0.8\linewidth]{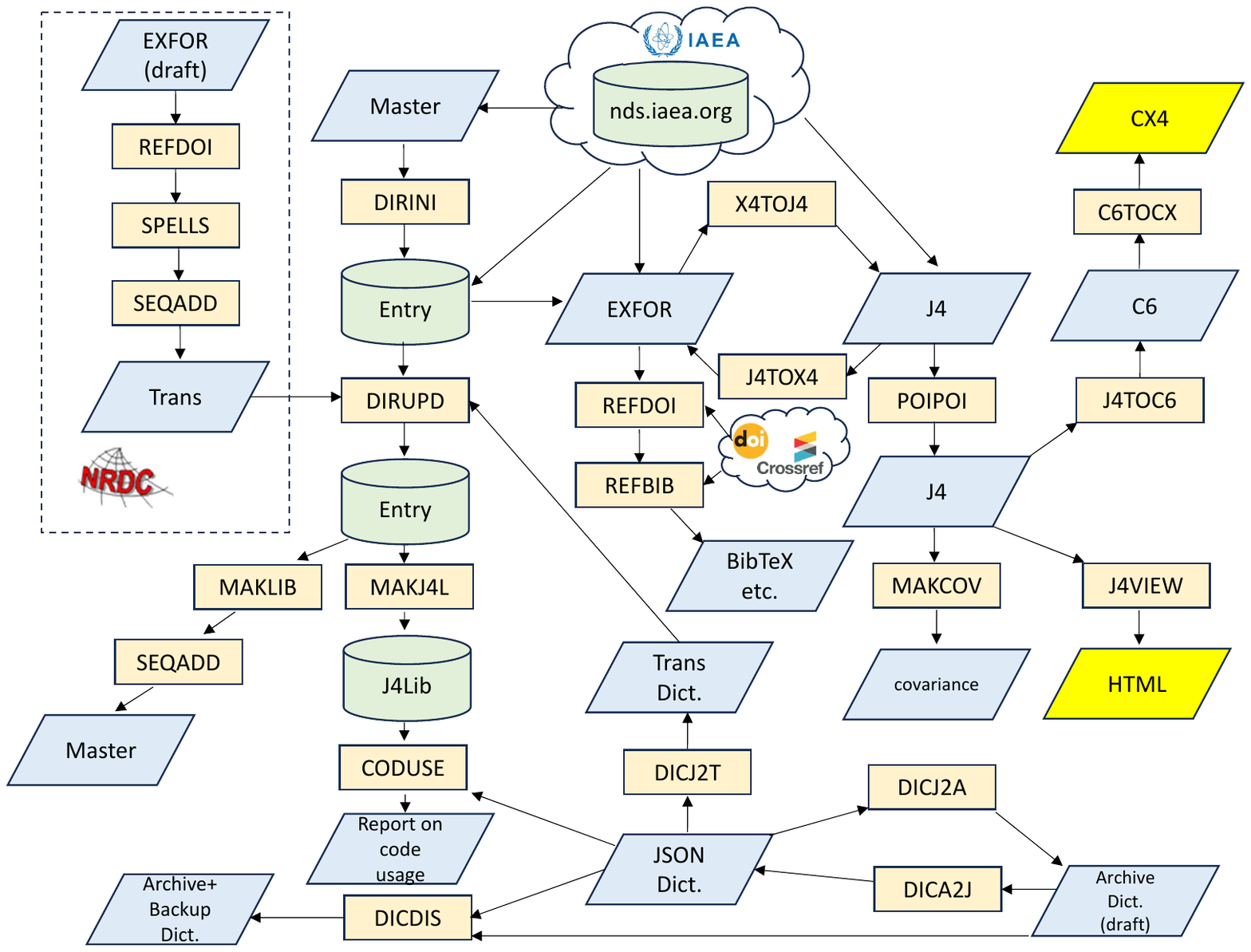}
\caption{
Flow of processing of EXFOR library and dictionary files by the ForEXy codes including generation of HTML and CX4 files.
``Master'', ``Entry'' and ``Trans'' are the abbreviations of the EXFOR Master File, EXFOR Entry Files and EXFOR Trans File, respectively.
}
\label{fig:forexyflow}
\end{figure}

The flow of the EXFOR file processing by X4VIEW (=X4TOJ4+POIPOI+J4VIEW) to obtain an HTML file can be traced on Fig.~\ref{fig:forexyflow}.

\section{X4TOCX: Converter for tabulation and plotting EXFOR data}
\label{sec:x4tocx}
We saw in Sect.~\ref{sec:evaluation} that sometimes users need to apply mathematical operations (e.g., conversion from isotopic cross section to elemental cross section, renormalization to the latest reference value) to the cross sections compiled in EXFOR to make the available experimental datasets more comparable.
For such a relatively simple correction and plotting,
cross section data in EXFOR files converted to a four-column ($E$, $\Delta E$, $\sigma$, $\Delta\sigma$) table format would be helpful.
X4TOCX is a Python module converting an EXFOR file to a CX4 (Compact EXFOR) file suitable for tabulation and plotting.
This script converts an EXFOR file to a CX4 file by calling four scripts: X4TOJ4 (EXFOR-to-J4 converter), POIPOI (post-processing of J4 to remove ``pointers" in EXFOR), J4TOC6 (J4-to-C6 converter) and C6TOCX (C6-to-CX4 converter).
 
Conversion of an EXFOR file \textit{exfor.txt} to a CX4 directory \textit{cx4} for the EXFOR entry D4083 can be done by
\begin{verbatim}
x4_x4tocx.py -i exfor.txt -e d4083 -o cx4
\end{verbatim}
.
The CX4 format does not support all but basic quantities defined in the EXFOR system.
For example,
the quantities not included in the MF scheme of the ENDF-6 format~\cite{Brown2023} (e.g., triple differential cross section, analyzing power) do not have their CX4 representations.
Construction of an experimental covariance matrix from an EXFOR file is sometimes required in data evaluation.
This is beyond the scope of tabulation by X4TOCX, but another ForEXy module MAKCOV can generate a covariance matrix from a J4 file as shown in Fig.~\ref{fig:forexyflow},
and it is used for simultaneous evaluation of fast neutron fission cross sections in the JENDL project~\cite{Otuka2025,Otuka2022}.

Figures~\ref{fig:CupxZnX4} and \ref{fig:CupxZnCX4} show an excerption of an EXFOR file providing $^\mathrm{nat}$Cu(p,x)$^{62,63,65}$Zn cross section datasets in an EXFOR entry and its conversion to CX4 for the $^{65}$Zn production part,
respectively.
The EXFOR file provides the datasets of three production cross sections together under the pointer 1 ($^{63}$Zn production), 2 ($^{62}$Zn production) and 3 ($^{65}$Zn production),
and it is not very clear which pieces of the DATA section are relevant to the $^{65}$Zn production.
During generation of the CX4 file for $^{65}$Zn production,
POIPOI reads the J4 file generated by X4TOC4, extracts the information relevant to the $^{65}$Zn production cross sections (i.e., the pieces provided without a pointer or with pointer 3),
and prints a new J4 file dedicated to $^{65}$Zn production.
The new J4 file is processed by J4TOC6 and C6TOCX to generate the CX4 file shown in Fig.~\ref{fig:CupxZnCX4}.
Cross section tables in CX4 files always consist of four data fields (laboratory incident energy $E$ and its uncertainty $\Delta E$ in eV, cross section $\sigma$ and its uncertainty $\Delta \sigma$ in barn).

\begin{figure}[hbtp]
\begin{center}
\begin{minipage}[c]{0.80\textwidth}
\begin{small}
\begin{verbatim}
SUBENT        D4083003   20160518                             D104D408300300001
BIB                  5         28                                 D408300300002
REACTION  1(29-CU-0(P,X)30-ZN-63,,SIG)                            D408300300003
          2(29-CU-0(P,X)30-ZN-62,,SIG)                            D408300300004
          3(29-CU-0(P,X)30-ZN-65,,SIG)                            D408300300005
...
DATA                 8         18                                 D408300300037
EN         EN-ERR     DATA      1ERR-T     1DATA      2ERR-T     2D408300300038
DATA      3ERR-T     3                                            D408300300039
MEV        MEV        MB         MB         MB         MB         D408300300040
MB         MB                                                     D408300300041
 5.7        0.4        121.       16.                             D408300300042
 103.       12.                                                   D408300300043
 6.5        0.4        164.       21.                             D408300300044
 136.       16.                                                   D408300300045
 7.2        0.4        187.       24.                             D408300300046
 152.       17.                                                   D408300300047
 8.2        0.4        245.       30.                             D408300300048
...
\end{verbatim}
\end{small}
\end{minipage}
\caption{EXFOR file providing $^\mathrm{nat}$Cu(p,x)$^{62,63,65}$Zn cross section datasets.}
\label{fig:CupxZnX4}
\end{center}
\end{figure}
\begin{figure}[hbtp]
\begin{center}
\begin{minipage}[c]{0.80\textwidth}
\begin{small}
\begin{verbatim}
# EXFOR #         : D4083.003.3
...
# Quantity        : natCu(p,x)65Zn independent production cross section
...
#---------------------------------------------------
# Einc        dEinc       sig         dsig        
#  eV          eV          b           b          
#---------------------------------------------------
  5.7000E+06  4.0000E+05  1.0300E-01  1.2000E-02
  6.5000E+06  4.0000E+05  1.3600E-01  1.6000E-02
  7.2000E+06  4.0000E+05  1.5200E-01  1.7000E-02
...
\end{verbatim}
\end{small}
\end{minipage}
\caption{CX4 file (preliminary) providing $^\mathrm{nat}$Cu(p,x)$^{65}$Zn cross section dataset converted from the EXFOR file shown in Fig.~\ref{fig:CupxZnX4}.}
\label{fig:CupxZnCX4}
\end{center}
\end{figure}

Figures~\ref{fig:17Opg18FX4} and \ref{fig:17Opg18FCX4} show an excerption of an EXFOR file providing two $^\mathrm{17}$O(p,$\gamma$)$^{18}$F cross section datasets in an EXFOR entry and its conversion to CX4 for a dataset from observation of secondary transitions~\footnote{Transitions other than those from the $^{18}$F excitation level initially formed following the proton capture (capture state).}.
The EXFOR file provides the astrophysical S-factors ($S$) as a function of the center-of-mass energy ($E_\mathrm{cm}$).
The center-of-mass energy is related to the laboratory incident energy $E_\mathrm{lab}$, target mass $m_1$ and projectile mass $m_2$ by
\begin{equation}
E_\mathrm{cm} = E_\mathrm{lab} \frac{m_1}{m_1+m_2},
\end{equation}
and the astrophysical S-factor is related to the cross section $\sigma$ and $E_\mathrm{cm}$ by
\begin{equation}
S=\sigma E_\mathrm{cm}\exp(2\pi\eta),
\end{equation}
where 
\begin{equation}
\eta=\alpha Z_1 Z_2 \sqrt{\frac{\mu}{2T_\mathrm{cm}}}
\end{equation}
with the fine structure constant $\alpha$, target and projectile atomic numbers $Z_1$ and $Z_2$, and reduced mass $\mu=m_1m_2/(m_1+m_2)$.
This equation shows that the conversion of astrophysical S-factors to cross sections is not very simple.
Figure~\ref{fig:17Opg18FCX4} shows that the $E_\mathrm{cm}$ and $S$ in the EXFOR file are converted to $E_\mathrm{lab}$ and $\sigma$ in the CX4 file,
which is more convenient for comparison with other datasets.
\begin{figure}[hbtp]
\begin{center}
\begin{minipage}[c]{0.80\textwidth}
\begin{small}
\begin{verbatim}
SUBENT        O2098002   20141022                             O053O209800200001
BIB                  6         22                                 O209800200002
REACTION  1(8-O-17(P,G)9-F-18,,SIG,,SFC)                          O209800200003
           S factors of primary transitions                       O209800200004
          2(8-O-17(P,G)9-F-18,,SIG,,SFC)                          O209800200005
           S factors of secondary transitions                     O209800200006
...
DATA                 5         11                                 O209800200034
EN-CM      DATA      1ERR-S     1DATA      2ERR-S     2           O209800200035
KEV        B*KEV      B*KEV      B*KEV      B*KEV                 O209800200036
 167.                             3.8        0.6                  O209800200037
 202.       6.4        0.6                                        O209800200038
 210.       6.1        0.7        5.6        0.5                  O209800200039
 228.       5.9        0.5        5.4        0.4                  O209800200040
...
\end{verbatim}
\end{small}
\end{minipage}
\caption{EXFOR file providing a $^{17}$O(p,$\gamma$)$^{18}$F astrophysical S-factor datasets.}
\label{fig:17Opg18FX4}
\end{center}
\end{figure}
\begin{figure}[hbtp]
\begin{center}
\begin{minipage}[c]{0.80\textwidth}
\begin{small}
\begin{verbatim}
# EXFOR #         : O2098.002.2
...
# Quantity        : 17O(p,g)18F cross section
...
#---------------------------------------------------
# Einc        dEinc       sig         dsig        
#  eV          eV          b           b          
#---------------------------------------------------
  1.7690E+05  0.0000E+00  1.4170E-10  2.2370E-11
  2.2245E+05  0.0000E+00  1.2834E-09  1.1460E-10
  2.4152E+05  0.0000E+00  2.2472E-09  1.6650E-10
...
\end{verbatim}
\end{small}
\end{minipage}
\caption{CX4 file (preliminary) providing $^{17}$O(p,$\gamma$)$^{18}$F cross section dataset converted from the EXFOR file shown in Fig.~\ref{fig:17Opg18FX4}.}
\label{fig:17Opg18FCX4}
\end{center}
\end{figure}

X4TOCX is under development,
and the usage and output of the code shown above may be different when the code is released as a part of the ForEXy code package.
Nevertheless,
the preliminary CX4 files generated from the latest EXFOR Entry File are regularly updated and distributed from the NRDC website~\cite{NRDCCX4} and NRDC GitHub repository~\cite{NRDCGit}.
We measured the CPU time and peak memory size to estimate the resource required for conversion of the whole EXFOR Entry Files (ver.~2026-08-20) to the CX4 files by using a computer equipped with an Intel Core i7-9700 CPU.
The results are $\sim$6000~s and $\sim$1~GB for conversion from EXFOR to J4 (without generation of DOIs for EXFOR references), $\sim$8000~s and $\sim$450~MB for conversion from J4 to C6, and $\sim$1000~s and $\sim$25~MB for conversion from C6 to CX4.
Each EXFOR entry is processed independently,
and we plan to reduce the time required for processing by parallelizing these conversion tools.
C6 files of about 20 EXFOR entries could not be converted to CX4 files during the full conversion due to compilation errors (e.g., presence of an outgoing angle or energy for cross sections integrated over the whole angle and energy range),
and the originating data centers are working to fix them.

\section{Summary}
\label{sec:summary}
The 80-column EXFOR format was designed as the internationally agreed format for the exchange of experimental nuclear reaction data between data centers.
The files in the EXFOR format are routinely created by and exchanged among the data centers.
We discussed challenges in use of the EXFOR files by EXFOR users such as evaluators of radionuclide production cross sections,
and introduced our attempts to convert an EXFOR file to (1) an HTML file for reading by humans (X4VIEW), and (2) CX4 four-column file for tabulation and plotting (X4TOCX).
One can freely download X4VIEW as a part of the code package ForEXy from the PyPI repository~\cite{PyPI2026} with its manual~\cite{Otuka2026}.
X4TOCX is still under development,
and we are working toward its release by extending the module to support additional output formats such as CSV.
We also plan to develop a new module plotting the datasets in the CX4 files,
and we expect the correctness of X4TOCX processing can be visually verified by systematic plotting by this new module.

\begin{acknowledgement}
We are grateful to S\'andor Tak\'acs for comments on the manuscript,
and to Alejandra Martinez and Arjan Koning for discussion.
The comments from Sophiya Taova and the experimentalists who participated in proofreading of the EXFOR drafts contributed to improvement of the X4VIEW module.
\end{acknowledgement}

\bibliography{radchem2026}

\begin{thebibliography}{10}

\bibitem{Otuka2015}
N.~Otuka and S.~Tak{\'a}cs.
\newblock Definitions of radioisotope thick target yields.
\newblock {\em Radiochimica Acta}, 103:1--6, 2015.

\bibitem{Okamoto1988}
K.~Okamoto.
\newblock {Proceedings of the IAEA Consultants' Meeting on Data Requirements
  for Medical Radioisotope Production, Tokyo, Japan, 20-24 April 1987}.
\newblock Technical Report INDC(NDS)-195, International Atomic Energy Agency,
  1988.

\bibitem{IAEA2001}
International Atomic~Energy Agency.
\newblock Charged particle cross-section database for medical radioisotope
  production: diagnostic radioisotopes and monitor reactions.
\newblock Technical Report INDC-TECDOC-1211, International Atomic Energy
  Agency, 2001.

\bibitem{Engle2019}
J.~W. Engle, A.~V. Ignatyuk, R.~Capote, B.~V. Carlson, A.~Hermanne, M.~A.
  Kellett, T.~Kib{\'e}di, G.~Kim, F.~G. Kondev, M.~Hussain, O.~Lebeda, A.~Luca,
  Y.~Nagai, H.~Naik, A.~L. Nichols, F.~M. Nortier, S.~V. Suryanarayana,
  S.~Tak{\'a}cs, F.~T. T{\'a}rk{\'a}nyi, and M.~Verpelli.
\newblock Recommended nuclear data for the production of selected therapeutic
  radionuclides.
\newblock {\em Nuclear Data Sheets}, 155:56--74, 2019.

\bibitem{Tarkanyi2019a}
F.~T. T{\'a}rk{\'a}nyi, A.~V. Ignatyuk, A.~Hermanne, R.~Capote, B.~V. Carlson,
  J.~W. Engle, M.~A. Kellett, T.~Kib{\'e}di, G.~N. Kim, F.~G. Kondev,
  M.~Hussain, O.~Lebeda, A.~Luca, Y.~Nagai, H.~Naik, A.~L. Nichols, F.~M.
  Nortier, S.~V. Suryanarayana, S.~Tak{\'a}cs, and M.~Verpelli.
\newblock Recommended nuclear data for medical radioisotope production:
  diagnostic positron emitters.
\newblock {\em Journal of Radioanalytical and Nuclear Chemistry}, 319:533--666,
  2019.

\bibitem{Tarkanyi2019b}
F.~T. T{\'a}rk{\'a}nyi, A.~V. Ignatyuk, A.~Hermanne, R.~Capote, B.~V. Carlson,
  J.~W. Engle, M.~A. Kellett, T.~Kibedi, G.~N. Kim, F.~G. Kondev, M.~Hussain,
  O.~Lebeda, A.~Luca, Y.~Nagai, H.~Naik, A.~L. Nichols, F.~M. Nortier, S.~V.
  Suryanarayana, S.~Tak{\'a}cs, and M.~Verpelli.
\newblock Recommended nuclear data for medical radioisotope production:
  diagnostic gamma emitters.
\newblock {\em Journal of Radioanalytical and Nuclear Chemistry}, 319:487--531,
  2019.

\bibitem{Hermanne2021}
A.~Hermanne, F.~T. T{\'a}rk{\'a}nyi, A.~V. Ignatyuk, S.~Tak{\'a}cs, and
  R.~Capote.
\newblock Upgrade of {IAEA} recommended data of selected nuclear reactions for
  production of {PET} and {SPECT} isotopes.
\newblock {\em Nuclear Data Sheets}, 173:285--308, 2021.

\bibitem{Tarkanyi2022}
F.~T{\'a}rk{\'a}nyi, A.~Hermanne, A.~V. Ignatyuk, S.~Tak{\'a}cs, and R.~Capote.
\newblock Upgrade of recommended nuclear cross section data base for production
  of therapeutic radionuclides.
\newblock {\em Journal of Radioanalytical and Nuclear Chemistry},
  331:1163--1206, 2022.

\bibitem{Hermanne2023a}
A.~Hermanne, F.~T. T{\'a}rk{\'a}nyi, A.~V. Ignatyuk, S.~Tak{\'a}cs, and
  R.~Capote.
\newblock Evaluated and recommended cross-section data for production of
  radionuclides with emerging interest in nuclear medicine imaging. part 1:
  Positron emission tomography ({PET}).
\newblock {\em Nuclear Instruments and Methods in Physics Research Section B:
  Beam Interactions with Materials and Atoms}, 535:149--192, 2023.

\bibitem{Hermanne2023b}
A.~Hermanne, F.~T. T{\'a}rk{\'a}nyi, A.~V. Ignatyuk, S.~Tak{\'a}cs, and
  R.~Capote.
\newblock Evaluated and recommended cross section data for production of
  radionuclides with emerging interest in nuclear medicine imaging. part 2:
  Single photon emission computed tomography ({SPECT}).
\newblock {\em Nuclear Instruments and Methods in Physics Research Section B:
  Beam Interactions with Materials and Atoms}, 544:165119, 2023.

\bibitem{Tarkanyi2024}
F.~T{\'a}rk{\'a}nyi, A.~Hermanne, A.~V. Ignatyuk, F.~Ditr{\'o}i, S.~Tak{\'a}cs,
  and R.~Capote~Noy.
\newblock Extension of recommended cross section database for production of
  therapeutic isotopes.
\newblock {\em Journal of Radioanalytical and Nuclear Chemistry}, 333:717--804,
  2024.

\bibitem{Hermanne2025}
A.~Hermanne, F.~T{\'a}rk{\'a}nyi, A.~V. Ignatyuk, S.~Tak{\'a}cs, and R.~Capote.
\newblock Critical re-evaluation of recommended excitation functions of 7
  reactions for $^{61}${Cu} formation.
\newblock {\em Nuclear Instruments and Methods in Physics Research Section B:
  Beam Interactions with Materials and Atoms}, 563:165690, 2025.

\bibitem{Otuka2014}
N.~Otuka, E.~Dupont, V.~Semkova, B.~Pritychenko, A.~I. Blokhin, M.~Aikawa,
  S.~Babykina, M.~Bossant, G.~Chen, S.~Dunaeva, R.~A. Forrest, T.~Fukahori,
  N.~Furutachi, S.~Ganesan, Z.~Ge, O.~O. Gritzay, M.~Herman, S.~Hlava\v{c},
  K.~Kat\={o}, B.~Lalremruata, Y.~O. Lee, A.~Makinaga, K.~Matsumoto,
  M.~Mikhaylyukova, G.~Pikulina, V.~G. Pronyaev, A.~Saxena, O.~Schwerer, S.~P.
  Simakov, N.~Soppera, R.~Suzuki, S.~Tak{\'a}cs, X.~Tao, S.~Taova,
  F.~T{\'a}rk{\'a}nyi, V.~V. Varlamov, J.~Wang, S.~C. Yang, V.~Zerkin, and
  Y.~Zhuang.
\newblock Towards a more complete and accurate experimental nuclear reaction
  data library ({EXFOR}): International collaboration between nuclear reaction
  data centres ({NRDC}).
\newblock {\em Nuclear Data Sheets}, 120:272--276, 2014.

\bibitem{LB}
H.~F. Schopper, editor.
\newblock {\em Production of radionuclides at intermediate energies}, volume~13
  of {\em Landolt-Boernstein numerical data and functional relationships in
  science and technology. Group I. Elementary particles, nuclei and atoms}.
\newblock Springer, 1991.

\bibitem{Pritychenko2011}
B.~Pritychenko, E.~B\v{e}t{\'a}k, M.~A. Kellett, B.~Singh, and J.~Totans.
\newblock The nuclear science references ({NSR}) database and web retrieval
  system.
\newblock {\em Nuclear Instruments and Methods in Physics Research Section A:
  Accelerators, Spectrometers, Detectors and Associated Equipment},
  640:213--218, 2011.

\bibitem{Pikulina2024}
G.~N. Pikulina and S.~M. Taova.
\newblock {EXFOR-Editor} package for entering, processing and representing
  nuclear reaction data in the {EXFOR} format.
\newblock {\em Journal of Nuclear Science and Technology}, 61:146--149, 2024.

\bibitem{Soppera2017}
N.~Soppera, M.~Bossant, O.~Cabellos, E.~Dupont, and C.~J. D{\'i}ez.
\newblock {JANIS: NEA JAva-based Nuclear Data Information System}.
\newblock {\em EPJ Web of Conferences}, 146:07006, 2017.

\bibitem{Forrest2014}
R.~A. Forrest, V.~Zerkin, and S.~Simakov.
\newblock Developments of the {EXFOR} database: Possible new formats.
\newblock {\em Nuclear Data Sheets}, 120:268--271, 2014.

\bibitem{Schnabel2020}
G.~Schnabel.
\newblock {A} computational {EXFOR} database.
\newblock {\em EPJ Web of Conferences}, 239:16001, 2020.

\bibitem{Otuka2025}
N.~Otuka, V.~Devi, and O.~Iwamoto.
\newblock {EXFOR} utility codes {(ForEXy)} and their application to neutron
  fission cross section evaluation.
\newblock {\em Applied Radiation and Isotopes}, 225:111903, 2025.

\bibitem{PyPI2026}
N.~Otuka.
\newblock {ForEXy: Utility codes for the EXFOR Library}.
\newblock https://pypi.org/project/forexy/.

\bibitem{Devi2025}
V.~Devi, N.~Otuka, and J.~S. Matharu.
\newblock Evaluation of $^{232}${Th} and $^{237}${Np} fast neutron-induced
  fission cross sections by simultaneous evaluation approach.
\newblock {\em EPJ Web of Conferences}, 322:10003, 2025.

\bibitem{Hermanne2018}
A.~Hermanne, A.~V. Ignatyuk, R.~Capote, B.~V. Carlson, J.~W. Engle, M.~A.
  Kellett, T.~Kib{\'e}di, G.~Kim, F.~G. Kondev, M.~Hussain, O.~Lebeda, A.~Luca,
  Y.~Nagai, H.~Naik, A.~L. Nichols, F.~M. Nortier, S.~V. Suryanarayana,
  S.~Tak{\'a}cs, F.~T. T{\'a}rk{\'a}nyi, and M.~Verpelli.
\newblock Reference cross sections for charged-particle monitor reactions.
\newblock {\em Nuclear Data Sheets}, 148:338--382, 2018.

\bibitem{Otuka2024}
N.~Otuka, S.~Tak{\'a}cs, M.~Aikawa, S.~Ebata, and H.~Haba.
\newblock Isomer production studied with simultaneous decay curve analysis for
  alpha-particle induced reactions on natural platinum up to 29~{MeV}.
\newblock {\em The European Physical Journal A}, 60:195, 2024.

\bibitem{Otuka2025b}
N.~Otuka, M.~Aikawa, S.~Tak{\'a}cs, D.~Gantumur, S.~Ebata, L.~Bold, A.~Nambu,
  and H.~Haba.
\newblock Energy dependence of isomeric ratios for alpha-particle-induced
  reactions on natural platinum up to 50~{MeV} studied by simultaneous decay
  curve analysis.
\newblock {\em The European Physical Journal A}, 61:184, 2025.

\bibitem{Zerkin2018}
V.~V. Zerkin and B.~Pritychenko.
\newblock The experimental nuclear reaction data ({EXFOR}): Extended computer
  database and web retrieval system.
\newblock {\em Nuclear Instruments and Methods in Physics Research Section A:
  Accelerators, Spectrometers, Detectors and Associated Equipment}, 888:31--43,
  2018.

\bibitem{Cullen2001}
D.~E. Cullen and A.~Trkov.
\newblock Program x4toc4 (version 2001-3): Translation of experimental data
  from the {EXFOR} format to a computation format.
\newblock Technical Report INDC-NDS-80, International Atomic Energy Agency,
  2001.

\bibitem{Koning2015}
A.~J. Koning.
\newblock {Bayesian Monte Carlo} method for nuclear data evaluation.
\newblock {\em The European Physical Journal A}, 51:184, 2015.

\bibitem{Kondev2021}
F.~G. Kondev, M.~Wang, W.~J. Huang, S.~Naimi, and G.~Audi.
\newblock The {NUBASE2020} evaluation of nuclear physics properties.
\newblock {\em Chinese Physics C}, 45:030001, 2021.

\bibitem{IAEA2017}
International Atomic~Energy Agency.
\newblock Cyclotron based production of technetium-99m.
\newblock Technical Report STI/PUB/1743, International Atomic Energy Agency,
  2017.

\bibitem{Meija2016}
J.~Meija, T.~B. Coplen, M.~Berglund, W.~A. Brand, P.~De~Bi{\`e}vre,
  M.~Gr\"{o}ning, N.~E. Holden, J.~Irrgeher, R.~D. Loss, T.~Walczyk, and
  T.~Prohaska.
\newblock Atomic weights of the elements 2013 ({IUPAC} technical report).
\newblock {\em Pure and Applied Chemistry}, 88:265--291, 2016.

\bibitem{Verpelli2011}
M.~Verpelli and D.~Abriola.
\newblock Information management tools for {Evaluated Nuclear Structure Data
  File} ({ENSDF}) interrogation and dissemination.
\newblock {\em Journal of the Korean Physical Society}, 59:1322--1324, 2011.

\bibitem{Levkovskij1991}
V.~N. Levkovskij, editor.
\newblock {\em Middle Mass Nuclides (A=40-100) activation cross sections by
  medium energy (E=10-50~MeV) protons and $\alpha$-particles (experiments and
  systematics)}.
\newblock INTER-VESTI, 1991.

\bibitem{Takacs2016}
S.~Tak{\'a}cs, F.~Ditr{\'o}i, M.~Aikawa, H.~Haba, and N.~Otuka.
\newblock Benchmark experiment for the cross section of the
  $^{100}${Mo}(p,2n)$^{99m}${Tc} and $^{100}${Mo}(p,pn)$^{99}${Mo} reactions.
\newblock {\em Nuclear Instruments and Methods in Physics Research Section B:
  Beam Interactions with Materials and Atoms}, 375:60--66, 2016.

\bibitem{Otuka2025a}
N.~Otuka.
\newblock {EXFOR Formats Manual}.
\newblock Technical Report IAEA-NDS-207 Rev. 2025/05, International Atomic
  Energy Agency, 2025.

\bibitem{Brown2023}
D.~A. Brown.
\newblock {ENDF-6} formats manual.
\newblock Technical Report BNL-224854-2023-INRE, Brookhaven National
  Laboratory, 2023.

\bibitem{Otuka2022}
N.~Otuka and O.~Iwamoto.
\newblock {EXFOR}-based simultaneous evaluation of neutron-induced uranium and
  plutonium fission cross sections for {JENDL-5}: {I}nputs and outputs.
\newblock Technical Report JAEA-Data/Code 2022-005, Japan Atomic Energy Agency,
  2022.

\bibitem{NRDCCX4}
{EXFOR Plot File}.
\newblock https://www.nds.iaea.org/nrdc/exfor-master/cx4/.

\bibitem{NRDCGit}
{EXFOR Plot File}.
\newblock https://github.com/iaea-nrdcnetwork/exfor-plot-file/,
  https://github.com/iaea-nds/exfor-plot-file/.

\bibitem{Otuka2026}
N.~Otuka.
\newblock {ForEXy: Utility codes for EXFOR}.
\newblock Technical Report IAEA-NDS-244 Rev. 2026/05, International Atomic
  Energy Agency, 2026.

\end{thebibliography}

\noindent
\textbf{Funding}\\
Not applicable

\end{document}